\documentclass[twocolumn,aps,pra,groupedaddress,amsmath,amssymb]{revtex4}
\usepackage{epsfig,amsmath,graphicx,amssymb}
\usepackage[usenames]{color}
\def\be{\begin{equation}}
\def\ee{\end{equation}}
\def\bea{\begin{eqnarray}}
\def\eea{\end{eqnarray}}
\def\bes{\begin{subequations}}
\def\ees{\end{subequations}}
\begin{document}

%%%%%%%%%%%%%%%%%%%%%%%%%%%%%%%%%%%%%%%%%%%%%%%%%%%%%%%%%%%%%%%%%%%%%%%%%%%%%%%%%
\title{Soliton Diffusion in a Bose-Einstein Condensate:  \\ A Signature of the Analogue Hawking Radiation}

\author{Chao Hang$^{1,3}$,  Gregory Gabadadze$^{2,3}$, and Guoxiang Huang$^{1,3}$}
\affiliation{$^1$State Key Laboratory of Precision Spectroscopy, School of Physical and Material Sciences,
                 East China Normal University, Shanghai 200062, China\\
$^2$Department of Physics, New York University, New York, NY 10003, USA\\
$^3$NYU-ECNU Institute of Physics at NYU Shanghai, Shanghai 200062, China
}
\date{\today}

\begin{abstract}

The analog Hawking radiation (HR) emanating from a sonic black hole horizon formed
in a cigar-shaped Bose-Einstein condensate (BEC) at finite temperature is investigated.
In particular, we study the effect of HR on a dark (topological) soliton.  We show that, due to thermal fluctuations,
the dark soliton in the BEC displays a nonlinear Brownian motion resulting in a specific asymmetric diffusion.
Based on numerical simulations we argue that the analog HR formed in the BEC can be detected indirectly
through measurement of the dark soliton diffusion.

\end{abstract}


\maketitle

{\it Introduction. } The goal of this work is to propose  a
scheme for an indirect detection  of the analog Hawking radiation in a certain
quantum system, using a topological soliton that the same system supports.

Hawking radiation (HR) is a quantum effect in a
classical background of a black hole ~\cite{Hawking}. It has been a subject of a numerous
studies aimed at better understanding  of the fundamental  principles of quantum physics
in the presence of a black hole (BH) geometry  (see, e.g., \cite{Marolf} and references therein).

In a commonly discussed astrophysical setting -- i.e., for a BH in the center of a galaxy --
the intensity of HR is negligible as compared to that of the in/out-flows
triggered by accretion of surrounding  matter and radiation onto the BH.
Hence, detection of HR in such settings  appears to be unlikely.

It is then desirable to look for different setups in which similar quantum radiation,
in otherwise classical background, might be expected and detected. Such a program had been
initiated by Unruh  who noticed the equivalence between the equations used to predict HR
and those governing wave propagation in certain inhomogeneous and moving media~\cite{Unruh0}.
The radiation of quantum counterparts of such waves, potentially
accessible in a lab, is referred as the analog HR.

A numerous interesting platforms  have been proposed for the detection of the analogue HR~\cite{Jacobson,Leonhardt0,Leonhardt1,Unruh1,Schutzhold,Philbin,Elazar,Garay,Wuster,Jain,Carusotto,Balbinot,Recati,Larre,Solnyshkov,Gerace,Giovanazzi,Horstmann,Belgiorno,Rousseaux,Weinfurtner1}. Among them,
atomic Bose-Einstein condensates (BECs) were suggested as potentially useful  for the
explorations. Moreover, a direct observation of the analog HR has  recently been claimed in such systems~
\cite{Lahav,Steinhauer0,Boiron,Steinhauer1}.

Given the complexity of such experiments and subtleties of the detection \cite{Lahav,Steinhauer0,Boiron,Steinhauer1},
it is highly desirable to have complementary methods of observing effects of  HR in BEC's.
For  instance,  indirect detection by observing  influence of HR on object within BEC
that can be controlled and manipulated would be welcome. In
this work we propose one such scheme using peculiarities of
diffusion of a topological (dark) soliton in BEC caused by a thermal radiation, the analog
HR in our case.

Solitons are fascinating large-amplitude excitations, stabilized by the interplay between nonlinearity and
dispersion/diffraction. They appear in nonlinear media or in nonlinear relativistic systems.
Among diverse types of solitons discovered in many branches of physics, matter wave
solitons in atomic  BECs have been  broadly studied~
\cite{Burger,Denschlag,Khaykovich,Strecker,Stellmer,Nguyen,
Fedichev,Muryshev,Sinha,Jackson,Cockburn,Efimkin0}.  Their  diffusion
has also been explored recently in various settings~\cite{McDonald,Efimkin,Aycock,Hurst}.

In this work we use the accumulated knowledge on the analog HR, on  solitons, and on
field theory techniques describing both of these phenomena,
to investigate the HR driven diffusion of dark (topological)
solitons in a cigar-shaped atomic BEC with a repulsive atom-atom interaction.

By developing an analytic approach based on effective field theory methods,
we show that a dark soliton immersed in BEC displays a
nonlinear temporal variance of its position under
the influence of thermal fluctuations,
corresponding to a nonlinear Brownian motion. Thus,
it exhibits an apparent, asymmetric diffusion during its propagation.
In particular, we find that the width and depth of the dark soliton change in time $t$ respectively
as,   $t^{3/2}\,T$ and $t^{-3/2}/T$,
with $T$ being temperature of thermal radiation.
Using these results, we propose a scheme for an indirect detection of  HR by measuring the diffusion
of the soliton caused by  the analog
HR emanating from a sonic black hole horizon in a moving BEC.

The proposed scheme  is entirely based on the inherent nonlinearity of  the BEC, and by that, it
differs  from previous studies. Since the dark soliton can be regarded as a large classical particle,
its diffusion  should be easier to detected than it is to detect the Hawking quanta directly.

\vskip 0.05in
%%%%%%%%%%%%%%%%%%%%%%%%%%
%

{\it The Model.--} We consider a cigar-shaped atomic BEC with repulsive atom-atom interaction that is
trapped by the potential $V_{\rm ext}(\mathbf{r})=\frac{M}{2}[\omega_{\bot}^2(x^2+y^2)+\omega_z^2z^2]$,
where $M$ is the atomic mass and $\omega_\bot$ ($\omega_z$) is transverse (axial) harmonic oscillator
frequency, with $\omega_{\perp}\gg \omega_z$. At  low energies and momenta the dynamics of such BEC
can be described by an effective Lagrangian for
a space-time dependent order parameter of the condensed state, $\psi(z,t)$, that can be referred as
a wave-function of the condensate.  In the long-wavelength approximation
both the condensate  and its fluctuations  are captured  by a classical solution
for $\psi(z,t)$ and its perturbations.  Below we will directly deal with the equation of motion  that follows
from the effective Lagrangian,  and also incorporates  the terms due to the external harmonic trap.
Thus, the wave-function $\psi(z,t)$,  satisfies  the Gross-Pitaevskii (GP) equation:
$$i\hbar\frac{\partial \psi}{\partial t}=\left(-\frac{\hbar^2}{2M}\frac{\partial^2}{\partial z^2}+\frac{M}{2}\omega_z^2z^2+\hbar\omega_\bot+g_{\rm{1D}}|\psi|^2\right)\psi,$$
where $g_{\rm{1D}}\equiv 2\hbar\omega_{\perp}a_s$, with $a_s>0$ is the $s$-wave scattering length determined  by
the quartic repulsive self-interaction term in the effective Lagrangian~\cite{Pethick}.
The condensate solution and spectrum of its long-wavelength perturbations ( the Bogliubov spectrum of
phonons) are well known;  we just emphasize for further use how  that spectrum changes
when one looks at a background flow of BEC with velocity $v_0$ along the $z$ direction
$$\omega=v_0 k\pm \left[\frac{k^2}{2M}\left(\frac{\hbar^2k^2}{2M}+2g_{\rm{1D}}\rho_0\right)\right]^{1/2},$$
where ``+'' (``$-$'') corresponds to the phonon propagating in the  $+z$ ($-z$) direction, while
$\rho_0$ is the one-dimensional density. The sound phase velocity at low momentum reads $(\omega/k)_{k=0}=v_0+c_s$,
where $c_s=\pm\sqrt{g_{\rm 1D}\rho_0/M}$ is the sound speed in the BEC frame.

The above description does not capture the effects of finite temperature.  In particular,
at any nonzero temperature a finite fraction of the atoms will not be in the condensed state and
will scatter incoherently off the condensed atoms, leading to dissipation in the condensate. Furthermore,
thermal fluctuations will give rise to randomness. To account for both of these effects
in a parametric way  one  modifies the GP equation
into a stochastic GP equation (SGPE)~\cite{Stoof,Gardiner}
\bea
i\hbar\frac{\partial \phi}{\partial
t}&=&(1-i\gamma)\left(-\frac{\hbar^2}{2M}\frac{\partial^2}{\partial z^2}+\frac{M}{2}\omega_z^2z^2+\hbar\omega_\bot\right.\notag\\
&&\left.+g_{\rm{1D}}|\phi|^2\right)\phi+\eta(z,t),
\label{SGPE} \eea
where $\gamma\equiv i\beta\hbar\Sigma^{K}(z,t)/4$ is the dissipation rate (with $\beta=1/(k_BT)$ and $k_B$ the Boltzmann constant), and $\hbar\Sigma^{K}$ is the Keldysh self-energy due to the incoherent collisions between condensed and noncondensed atoms~\cite{note000}. The last term $\eta$ parametrizes  the
thermal fluctuation, obeying the fluctuation-dissipation relation
$$\langle\eta(z,t)\eta^\ast(z',t')\rangle=2\hbar k_B T \gamma(z,t)\delta(z-z')\delta(t-t'),$$
with $\langle\cdots\rangle$ denoting the averaging over different noise realizations.
Since the BEC is highly elongated and the region of interest is far away from the edges,
an approximate space- and time-independent dissipation rate can be used in the analytical treatment,
i.e., we  adopt the approximation $\gamma\approx\gamma(0)=3Mk_BTa_s^2/(\pi\hbar^2)$~\cite{Penckwitt}.

For convenience of later calculations, we rewrite the SGPE (\ref{SGPE}) into the dimensionless form
\bea i\frac{\partial F}{\partial\tau}&=&(1-i\gamma)\left(1-\tilde{\mu}-i\tilde{k}_0\frac{\partial}{\partial
\zeta}-\frac{1}{2}\frac{\partial^2 }{\partial
\zeta^2}+\frac{\Omega^2}{2}\zeta^2\right.\notag\\
&&\left.+g|F|^2\right)F +\Lambda, \label{SGPE1} \eea
where $\tau=\omega_{\bot}t$, $\zeta=l_{\bot}^{-1}z$, $g=2a_s\rho_0$, $F=\phi e^{i\tilde{\mu}\tau-i\tilde{k}_0\zeta+i\tilde{\omega}_0\tau}/\sqrt{\rho_0}$, $\Omega=\omega_z/\omega_\perp\ll1$, $\tilde{\mu}=\mu/(\hbar\omega_{\bot})$, and
$\Lambda=\eta e^{i\tilde{\mu}\tau-i\tilde{k}_0\zeta+i\tilde{\omega}_0\tau}/(\hbar\omega_{\bot})$, with $l_{\perp}\equiv \sqrt{\hbar/(M\omega_{\perp})}$ being the transverse harmonic-oscillator length, $\tilde{k}_0= k_0l_{\bot}$ being the flow wavenumber, and $\tilde{\omega}_0=\omega_0/\omega_{\bot}$ [$\omega_0\equiv Mv_0^2/(2\hbar)$] the flow frequency. The correlator of the dimensionless fluctuation fields  reads $\langle\Lambda^\ast(\zeta,\tau)\Lambda(\zeta',\tau')\rangle=2 k_B T\gamma\delta(\zeta-\zeta')\delta(\tau-\tau')/(\hbar\omega_{\bot}^2)$.

If $\gamma$, $\Omega$, $\Lambda$ vanish, then the dimensionless SGPE (\ref{SGPE1}) admits an
exact dark soliton solution~\cite{Frantz}
$$F=\sqrt{(\tilde{\mu}-1)/g}(\cos\phi_0\tanh\tilde{Z}+i\sin\phi_0),$$ with $\tilde{Z}=\sqrt{\tilde{\mu}-1}\cos\phi_0[\zeta-(\tilde{k}_0+\sqrt{\tilde{\mu}-1}\sin\phi_0)\tau-\zeta_0].$
Here, $\zeta_0$ denotes the initial position of the soliton center and $\phi_0$ ($|\phi_0|<\pi/2$) is a phase
characterizing the blackness (the difference between the minimum soliton intensity and the
background intensity)  as $[(\tilde{\mu}-1)/g]\cos^2\phi_0$, its width  $3.3/(\sqrt{\tilde{\mu}-1}\cos\phi_0)$, and  its
velocity $\tilde{k}_0+\sqrt{\tilde{\mu}-1}\sin\phi_0$.

In the next section we will consider small but nonzero $\gamma$, $\Omega$,  and $\Lambda$,
and their effects on the soliton.

\vskip 0.05in

{\it Diffusion of the soliton.--} We start by looking  for analytical results on soliton diffusion,
to get a clear  physical picture and then to  compare with numerical simulations.  Starting with the
SGPE (\ref {SGPE}) in a weak nonlinear regime
a stochastic Korteweg-de Vries (KdV) equation  can be  derived by using the the multi-scale
method developed in Ref.~\cite{Huang}; the result reads as follows:
\be
\frac{\partial u}{\partial \tau}+\frac{3\tilde{c}_s}{u_0}u\frac{\partial u}{\partial \tilde{\zeta}}-\frac{1}{8\tilde{c}_s}\frac{\partial^3 u}{\partial \tilde{\zeta}^3}=-\frac{1}{4\tilde{c}_s}\frac{\partial \Lambda}{\partial \tilde{\zeta}}-\mathcal{R}(u,\gamma), \label{kdv1} \ee
where $\tilde{\zeta}=\zeta-(\tilde{k}_0+\tilde{c}_s)\tau$ and $u=F\exp(-i\varphi)-u_0$~\cite{Supplementary}. The first term on the right hand side of Eq.~(\ref{kdv1}) results from the thermal fluctuation; the second term, $\mathcal{R}$, comes from the higher-order effects  -- e.g., the
high-order thermal fluctuations, higher-order dispersion and nonlinearity, the trapping potential along the $z$ direction, etc. --
which are all small and neglected in the analytical treatment,
but will be included in our numerical simulations.

If $\Lambda$ is set to zero, then Eq.~(\ref{kdv1}) admits an exact dark soliton solution
$u=-A_0{\rm sech}^2[\sqrt{2\tilde{c}_s^{2}A_0/u_0}(\tilde{\zeta}+\tilde{c}_s
A_0\tau/u_0-\zeta_0)]$, where $A_0$ is a positive constant characterizing the amplitude of the soliton.
In the leading-order approximation the dark soliton solution of Eq.~(\ref{kdv1}) can be expressed as
$$F(\zeta,\tau)=u_0\left[1-\tilde{A}_0{\rm sech}^2\left(\sqrt{2\tilde{c}_s^{2}\tilde{A}_0}X\right)\right]
e^{i\varphi},$$
%\label{dark_soliton1}
with $\tilde{A}_0=A_0/u_0$, $X=\zeta-[\tilde{k}_0+\tilde{c}_s(1-\tilde{A}_0)]\tau-\zeta_0$, and
$\varphi=-\sqrt{2\tilde{A}_0}\tanh(\sqrt{2\tilde{c}_s^{2}\tilde{A}_0}X)$. We should note that:
 (i)~only the small-amplitude dark solitons can be described by the KdV equation;
 (ii)~dark solitons can be at rest in the lab frame if $\tilde{k}_0+\tilde{c}_s(1-\tilde{A}_0)=0$,
 which can only be satisfied if there is a background flow (i.e. $\tilde{k}_0\neq 0$).

If the thermal fluctuations are weak enough ($\Lambda\ll 1$), the first term on the right hand side of Eq.~(\ref{kdv1}) can be treated as a small perturbation. Thus, its effect on the dark soliton can be studied by perturbing the non-perturbative soliton solution ~\cite{Herman}; this leads to the  following  perturbed solution
\bea
u&=&-(A_0+\sqrt{2A_0}W_1){\rm sech}^2\left[\left(\sqrt{2\tilde{c}_s^2\tilde{A}_0}
+\frac{\tilde{c}_s}{\sqrt{u_0}}W_1\right)X\right.\nonumber\\
&&\left.+W_2\right],\label{ds}
\eea
with $W_1=[1/(2\sqrt{u_0})]\int_0^{\tau}d\tau'\int_{-\infty}^\infty ds\, \Lambda \tanh s\,{\rm sech}^2s$, and
$W_2=(\tilde{c}_s^2\tilde{A}_0/u_0)\int_0^{\tau}d\tau'
\int_0^{\tau'}d\tau''\int_{-\infty}^\infty ds\, \Lambda \tanh s\,{\rm sech}^2s-[1/(2u_0\sqrt{2\tilde{A}_0})]\int_0^{\tau}d\tau'\int_{-\infty}^\infty ds\, \Lambda (s\sinh s-\cosh s)/\cosh^3 s$. %From the solution~(\ref{ds}) one can see that under the influence of the random fluctuation, the blackness of the soliton changes from $\tilde{A}_0$ to $\tilde{A}_0+\sqrt{\frac{2\tilde{A}_0}{u_0}}W_1$ and its full width at half maximum (FWHM) changes from $\frac{1.76}{\sqrt{2\tilde{c}_s^2\tilde{A}_0}}$ to $\frac{1.76}{\left(\sqrt{2\tilde{c}_s^2\tilde{A}_0}+\frac{\tilde{c}_s}{\sqrt{u_0}}W_1\right)}$.

In what follows we'll be  interested in the long-time behavior of the dark soliton under the influence of the thermal fluctuations. To this end, we compute the ensemble average $\langle u \rangle$, and consider the case of $\tau\gg1$. Following the method of Ref.~\cite{Wadati}, and using the  relations $\langle W_1^2 \rangle =u_0^2\Lambda_0\sigma/(15\tilde{c}_s^4)$, $\langle W_1W_2 \rangle=2u_0^2A_0\Lambda_0\sigma^2/(15\tilde{c}_s^4)$, and $\langle W_2^2 \rangle=u_0^2\Lambda_0\sigma/(15\tilde{c}_s^4)[(30+\pi^2)/(24A_0)
+4A_0^2\sigma^2]$, we obtain the asymptotic solution for $\tau\gg1$
%~\cite{Supplementary}:
%
\bea
\langle u \rangle &=&-\frac{u_0^{5/4}}{3\sqrt{\pi}\tilde{c}_s}\sqrt{\frac{10}{3\Lambda_0\tau^3}}
\,\exp \left ( {-\frac{5\sqrt{u_0}}{3\tilde{A}_0\Lambda_0\tau^3}X^2} \right )\,,  \label{asymp}
\eea
where $\Lambda_0\equiv2 k_B T \gamma/(\hbar\omega_{\perp}^2)\propto T^2$. From Eq.~(\ref{asymp}) we see that for a long-time diffusion, the width of the dark soliton grows as  $t^{3/2}T$ while the blackness decreases as $1/ (t^{3/2}T)$. Even though  the soliton gets deformated because of the diffusion,  the  area enclosed by its shape-function remains  constant.

The obtained soliton diffusion is due to its Brownian motion induced by the thermal fluctuation in the BEC. From the solution~(\ref{ds}),
we can easily extract the time dependence of the position of the center of the soliton, $\zeta_{\rm DS}=\zeta_0+[\tilde{k}_0+\tilde{c}_s(1-\tilde{A}_0)]\tau
-\sqrt{u_0}W_2/[\tilde{c}_s(\sqrt{2A_0}+W_1)]$. From this, we can calculate the variance of the position, $D(\zeta_{\rm DS})\equiv\langle (\zeta_{\rm DS}-\langle \zeta_{\rm DS} \rangle)^2 \rangle$,  which reads
\be D(\zeta_{\rm DS})\approx\frac{1}{\sqrt{u_0}}\left(\frac{30+\pi^2}{45}\frac{1}{16\tilde{c}_s^4 \tilde{A}_0^2}+\frac{2}{15}\tilde{A}_0\tau^2\right)\Lambda_0\tau. \label{var}
\ee
We see that for small $\tau$, $D(\zeta_{\rm DS})\propto\tau$, in accordance with the Einstein relation for Brownian motion~\cite{Einstein}; however for larger $\tau$, $D(\zeta_{\rm DS})\propto\tau^3$, that is,  the dark soliton displays a nonlinear time-variance of its position, which can be regarded  as a nonlinear Brownian motion. Furthermore, in contrast with the Einstein relation where the diffusion coefficient $\propto1/\gamma$, here $D(\zeta_{\rm DS})\propto\gamma$ since  the dark soliton has a negative mass~\cite{Aycock,Hurst}.

Having the theoretical part clarified we adopt realistic parameters based on a recent experiment reported in Ref.~\cite{Stellmer},
where a cigar-shaped almost pure $^{87}$Rb BEC of $N\approx5\times10^4$ atoms in the $\langle 5^{2}S_{1/2},F=1,m_F=-1\rangle$ state was
prepared, with $(\omega_{\bot},\omega_{z})=2\pi\times\{133,5.9\}$ Hz and the peak density around $5.8\times10^{13}$ cm$^{-3}$.
Furthermore, the velocity of the background flow is taken to be $v_0\approx0.7$ mm s$^{-1}$, and the temperature of the BEC is assumed to be
$T\approx5$ nK (which is much lower than the critical temperature of one-dimensional BEC, $T_c=N\hbar\omega_z/[{\rm ln}(2N)k_B]\approx1.2$
$\mu$K~\cite{Ketterle}). With the above parameters at hand, one calculates $l_{\bot}\approx1$ $\mu$m, $c_{s}\approx\pm0.8$ mm s$^{-1}$, and $\gamma\approx0.22\times10^{-4}$.

\begin{figure}[htb] \centering
\includegraphics[width=0.96\columnwidth]{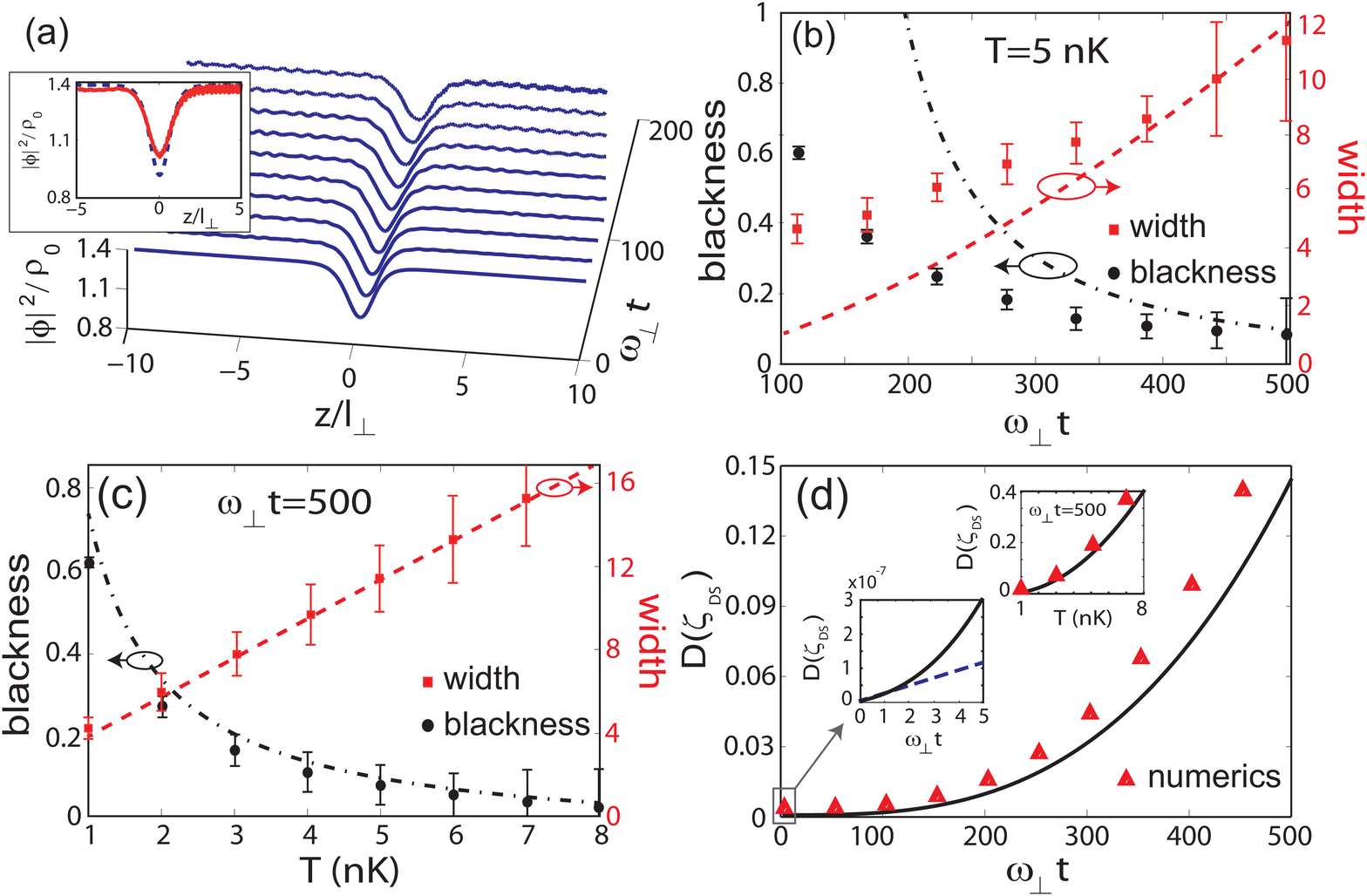}
\caption{(color online) (a): Dimensionless density $|\phi|^2/\rho_0$ as a function of $z/l_{\perp}$ and $\omega_{\perp} t$, obtained for the initial position $\zeta_0=0$, phase $\phi_0\approx-0.86$ rad, and the dissipation rate $\gamma\approx0.22\times10^{-4}$ (corresponding to the temperature $T\approx5$ nK) by numerically solving the SGPE. Inset: Intensity of the dark soliton at $t=0$ (blue dashed line) and $t\approx300$ ms (red solid line). (b) and (c): Dark soliton's width (red squares) and blackness (black circles) as functions of $\omega_{\bot}t$ for $T\approx5$ nK (b) and as functions of $T$ for $t\approx600$ ms (c). Error bars indicate the standard deviation from the mean value for 50 runs. The red dashed (black dashed-dotted) lines stand for the asymptotic behavior of the dark soliton's width (blackness) predicted by Eq.~(\ref{asymp}). (d): The variance of the position of the center of the dark soliton, $D(\zeta_{\rm DS})$, as a function of $\omega_{\perp} t$ obtained from the numerical simulation (red triangles) and from the analytic prediction~(\ref{var}) (black solid line). Left lower inset: $D(\zeta_{\rm DS})$ (black solid line) and its linear part (blue dashed line) in the range of $0<\omega_{\bot}t<5$. Right upper inset: The dependence of $\zeta_{\rm DS}$ on $T$ for $t\approx600$ ms. }\label{fig1}
\end{figure}

To test the obtained analytical results, we solve the SGPE  numerically by using the spectral method together with the
fourth-order Runge-Kutta method for time stepping. The numerical evolution of a typical dark soliton is shown in Fig.~\ref{fig1}(a), with the initial position and phase respectively taken as $\zeta_0=0$ and $\phi_0\approx-0.86$ rad (the initial blackness and width are approximately 0.6 and 4.3, respectively).
%===========================fig1===============================%
%===========================fig1===============================%
Fig.~\ref{fig1}(b) shows the width (red squares) and the blackness (black circles) of the dark soliton as functions of $\tau$ for $T\approx5$ nK, obtained by averaging over 50 runs of independent noise realizations. The respective analytic curves obtained  are also shown by the red dashed and black dashed-dotted lines, respectively. We see that the numerical results agree with the analytic ones for large $\tau$, however, a disagreement occurs for small $\tau$ since  the asymptotic solution (\ref{asymp}) is valid only in the long-time regime (i.e. $\tau\gg1$). Fig.~\ref{fig1}(c) shows the same quantities as functions of $T$ for $\tau=500$ ($t\approx600$ ms). Since $\tau\gg1$, the numerical results agree well with the analytic predictions in the entire range of $T$.

The dependence of $D(\zeta_{\rm DS})$ on $\tau$ is illustrated in Fig.~\ref{fig1}(d), where the red triangles are obtained from the numerical
calculation and the black solid line is  the analytic  result~(\ref{var}). One can see that the numerical and analytic results are matched quite well.
%A nonlinear-in-time variance appears for large $\tau$.
The left lower inset of the figure shows $D(\zeta_{\rm DS})$ (black solid line) and its linear part (blue dashed line) in the range of $0\leq\tau\leq5$. One sees that they are matched only for a very short time, $0<\tau\lesssim2$ ($0<t\lesssim2.4$ ms). Thus, for a long-time behavior, the dark soliton displays a nonlinear Brownian motion instead of a linear one. The dependence of $\zeta_{\rm DS}$ on $T$ for $\tau=500$ ($t\approx600$ ms) is also given, as shown in the right-upper inset.

\vskip 0.05in

\begin{figure}[htb] \centering
\includegraphics[width=0.96\columnwidth]{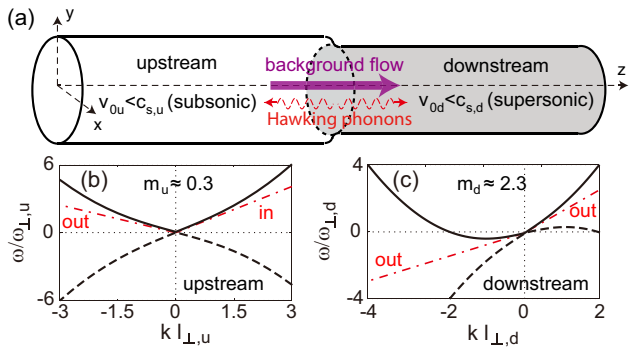}
\caption{(color online) (a): The setup scheme for mimicking sonic black hole, with notations defined in the text. (b) and (c): Bogoliubov spectrum of the linear excitations (phonons) on the background flow in the upstream (a) and downstream (b) of the BEC. The Mach number in the upstream (downstream) is $m_u=v_{0u}/|c_{s,u}|\approx0.3<1$ ($m_d=v_{0d}/|c_{s,d}|\approx2.3>1$). Red dashed-dotted lines denote the slope (sound speed) of long-wavelength phonon excitation near $k=0$. ``in'' and ``out'' denote in-going and out-going directions with respect to the black hole horizon, respectively. }\label{fig2}
\end{figure}

%%%%%%%%%%%%%%%%%%%%%%%%%%%%%%%%%%%%%%%%%%%%%%%%%%%%%%%%%%%%
{\it Detection on Analogue HR via soliton diffusion.--} We now turn to demonstrate that the measurement of the dark soliton diffusion might be an effective technique to detect the analogue HR in the BEC. To this end, we assume that the background flow in the BEC is generated along the $z$ direction; this can be realized by, say, adiabatically accelerating the trapping potential in the $z$ direction until the BEC reaches a constant velocity $v_0$~\cite{Lahav}.
The near-horizon geometry of an analogue sonic black hole can be mimicked by a transition region from a subsonic to supersonic background flow [see Fig.~\ref{fig2}(a)].
%===========================fig2===============================%
%===========================fig2===============================%
If the central position of the horizon is at $z=0$, the upstream (downstream) of the BEC corresponds to the region $z<0$ ($z>0$), where the flow is subsonic (supersonic). The analog HR is a thermal radiation of the phonons emanating from the analog horizon. Since the background flow
is supersonic  in the downstream region, and the  phonons cannot escape from it, that region can be regarded as the interior of an analog black hole.

In practice, to create a horizon of the analogue black hole, we assume there is a stationary, step-like change of the radial confinement of the trapping potential, i.e. $\omega_{\perp}=\omega_{\perp,u}$ for $z<0$ and $\omega_{\perp}=\omega_{\perp,d}$ for $z>0$, with $\omega_{\perp,u}<\omega_{\perp,d}$ (hereafter, subscripts ``$u$'' and ``$d$'' refer to the upstream and downstream, respectively).
Accordingly, the transverse harmonic-oscillator length in the upstream (downstream) is $l_{\perp,u}\equiv \sqrt{\hbar/(M\omega_{\perp,u})}$ ($l_{\perp,d}\equiv \sqrt{\hbar/(M\omega_{\perp,d})}$). The quantities in the upstream and  downstream regions can be related to  one another by the mass-current conservation and the condition of equal chemical potential in the whole BEC~\cite{Larre}: {$\rho_{0u} v_{0u} = \rho_{0d} v_{0d}$}, and $\frac{M}{2}v_{0u}^2+\hbar\omega_{\bot,u}+g_{{\rm 1D},u}\rho_{0u}=\frac{M}{2}v_{0d}^2+\hbar\omega_{\bot,d}+g_{{\rm 1D},d}\rho_{0d}$, respectively, with $g_{{\rm 1D},u}=2\hbar\omega_{\perp,u}a_{s}$ and $g_{{\rm 1D},d}=2\hbar\omega_{\perp,d}a_{s}$.

The sonic black hole requires that $v_{0u}<|c_{s,u}|$ and $v_{0d}>|c_{s,d}|$, where the sound
speeds in the upstream (downstream) is $c_{s,u}=\pm\sqrt{g_{{\rm 1D},u}\rho_{0u}/M}$ ($c_{s,d}=\pm\sqrt{g_{{\rm 1D},d}\rho_{0d}/M}$). For a simple case with $v_{0u}=|c_{s,u}|-u$ and $v_{0d}=|c_{s,d}|+u$ ($0<u\ll |c_{s,u}|,\,|c_{s,d}|$), the analogue Hawking temperature
is determined by the velocity gradient (see, e.g.,  ~\cite{Lahav})
$$T_H \approx \frac{\hbar}{2\pi k_B}|\frac{\partial}{\partial z}(v-c_s)|\approx \frac{\hbar u}{\pi k_B l_0},$$
where $l_0=\hbar/(M|c_s|)$ is the healing length. The latter is assumed to be the shortest length scale
in the hydrodynamical limit where the wavelength of the Bogoliubov excitations are much longer than $l_0$~\cite{note100}.

For the numerical calculations, we take $a_{s}=94.8\,a_0$, $\omega_{\bot,u}=2\pi\times133$ Hz ($l_{\bot,u}\approx0.94$ $\mu$m), {$\omega_{\bot,d}=2\pi\times140$ Hz ($l_{\bot,d}\approx0.91$ $\mu$m)}, $\rho_{0u}\approx1.0\times10^{6}$ cm$^{-1}$, $v_{0u}=0.25$ mm s$^{-1}$, and $c_{s,u}\approx\pm0.78$ mm s$^{-1}$. With the above parameters we get $\rho_{0d}\approx2.67\times10^{5}$ cm$^{-1}$, $v_{0d}\approx0.94$ mm s$^{-1}$, and $c_{s,d}\approx\pm0.41$ mm s$^{-1}$. Thus, the Mach number in the upstream (downstream) reads: $m_u=v_{0u}/|c_{s,u}|\approx0.3<1$ ($m_d=v_{0d}/|c_{s,d}|\approx2.3>1$), $u\approx0.5$ mm s$^{-1}$, and the Hawking temperature $T_H\approx1.4$ nK.

Fig.~\ref{fig2}(b) and Fig.~\ref{fig2}(c) show the Bogoliubov spectrum in the upstream and  downstream regions, respectively. We see that long-wavelength excitations in the upstream are able to propagate in both directions, i.e., they'd be in-going and out-going with respect to the horizon. However, they are dragged away by the background flow and are unable to propagate back to the black hole horizon in the downstream, i.e. they can only propagate along the out-going direction with respect to the horizon. The minimum wavelength of the trapped, long-wavelength excitations in the downstream is approximately 5.5 $\mu$m.

Since the system is essentially  one-dimensional, we assume that a uniform density region will quickly equilibrated at the Hawking temperature $T_H$, as long as the latter is higher than the ambient temperature of the BEC. Once this happens, a dark soliton will be injected in the upstream region, somewhat  close to the horizon, but in a way for it not to back-react on the analog horizon significantly; this can by done by employing the known phase-imprinting laser field~\cite{Stellmer}, allowing to precisely set the soliton  position, blackness, and velocity.

If $T_H$ is higher than the temperature of all other reservoirs that may exist, the dark soliton will behaves as a nonlinear Brownian particle due to the influence of the HR, as detailed in the previous section. Note that the coolest  BEC realized up to now can  have temperatures as low as
$\sim10$ pK~\cite{Kova}. In such setups the role of the BEC thermal fluctuations will  indeed be negligible and only the HR  will remain responsible for the soliton diffusion. Furthermore, since the dark soliton is set  with zero initial velocity [when taking $\tilde{A}_0=\tilde{k}_{0u}/\tilde{c}_{s,u}+1\approx0.24$ ($\tilde{c}_{s,u}<0$)] and a small amplitude, it is ``slow", and hence sensitive to the influence of the HR.

\begin{figure}[htb] \centering
\includegraphics[width=0.98\columnwidth]{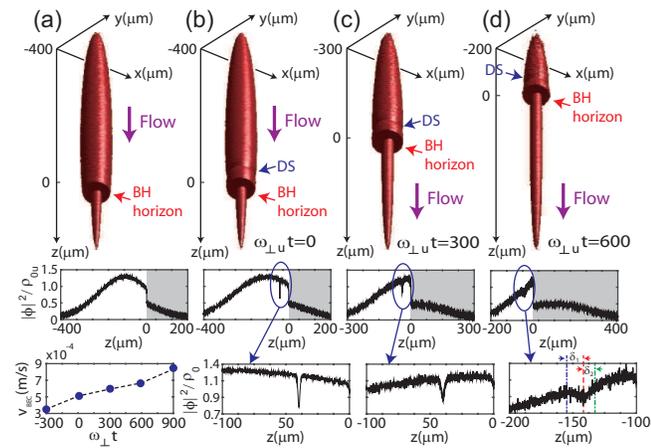}
\caption{(color online) The scheme for detecting the analogue HR in the BEC. (a): A cigar-shaped BEC (with length $\approx 500$ $\mu$m and flow velocity $v_0=0.25$ mm s$^{-1}$) passes over the change of radial confinement at $z=0$, allowing the occurrence of the black hole horizon and the analogue HR. (b): When the uniform density region of the BEC reaches the horizon  $t=0$, a dark soliton is generated in the upstream. (c) and (d): The diffusion of the dark soliton due to the influence of the HR at $\omega_{\perp,u}t=300$ ($t\approx359$ ms) (c) and 600 ($t\approx718$ ms) (d), respectively. Middle row: Dimensionless density $|\phi|^2/\rho_{0u}$ along the $z$ direction, corresponding to (a)-(d), where the white and shaded regions denote the upstream and the downstream, respectively. Lower row: The first panel shows the velocity of the whole BEC at $\omega_{\perp,u}t=-300$, 0, 300, 600, and 900 ($t\approx-359$, 0, 359, 718, and 1077 ms). The rest panels zoom in the dark solitons. $\delta_1$ ($\delta_2$) in the last panel denotes the distance between the soliton center and its left (right) border. }\label{fig3}
\end{figure}

Last but not least, we perform the numerical simulation of the  entire process of detection of the analogue HR. In the simulation, we first let the BEC (with the length $\approx500$ $\mu$m and a background flow velocity $v_0=0.25$ mm s$^{-1}$) to undergo  the change in the radial confinement at $z=0$, enabling
the emergence of the analogue black hole horizon  [Fig.~\ref{fig3}(a)].
%===========================fig3===============================%
%===========================fig3===============================%
As soon as  the uniform density region reaches the horizon, a dark soliton with zero initial velocity is injected in the upstream (we set this time as $t=0$) [Fig.~\ref{fig3}(b)]. Shown in Fig.~\ref{fig3}(c) and Fig.~\ref{fig3}(d) are numerical results on the HR-induced-diffusion of the dark soliton at $\tau\approx300$ ($t\approx359$ ms) and $\tau\approx600$ ($t\approx718$ ms), respectively. From Fig.~\ref{fig3}(c), we see that the soliton width (blackness) has increased (decreased) by nearly 1.5 times as compared with the initial values.  These changes   are in good agreement with the results given in Fig.~\ref{fig2}(b), and  should be possible to measure in a realistic experiment (see the third panel in the middle row of Fig.~\ref{fig3}). In Fig.~\ref{fig3}(d), an asymmetric soliton diffusion is observed. Specifically, we measure the distance between the soliton center and its left (right) "edge", denoted by $\delta_1$ ($\delta_2$). We find that $\delta_1/\delta_2\approx1.3>1$ (the last panel in the lower row of Fig.~\ref{fig3}). {Such an asymmetric soliton diffusion might be caused by the inhomogeneous density of the BEC along the longitudinal direction or by an asymmetry in the pressure acting on the soliton due to the HR.} Furthermore, we calculate the velocity of the whole BEC (i.e. the velocity of the center-of-mass of the BEC), $v_{\rm BEC}$, at different times,  and  the influence of the HR is also calculated (see the first panel of the lower row in Fig.~\ref{fig3}). One can see that $v_{\rm BEC}$ increases from the value of the background flow velocity in the upstream to that in the downstream. Our simulation confirms  that the dark soliton propagation in BEC is robust, and  the soliton has  the lifetime longer than 718 ms.

{\it Conclusion.--}  The Hawking radiation is one of the intriguing long studied quantum phenomenon in
a  classical background field of a black hole. Its detection in any  realistic astrophysical context is  unlikely. The analog Hawking radiation
is governed by the mathematical equations that are identical to those used in the Hawking  calculation. Therefore,
it is interesting to study the analog HR in table-top experimental systems where it can be potentially detected directly or indirectly.

In this work the analog HR emanating from a sonic black hole horizon formed
in a cigar-shaped Bose-Einstein condensate was investigated.
In particular, we studied the effect of HR on a dark (topological) soliton.  We showed that under the thermal HR
the dark soliton in the BEC  would display a nonlinear Brownian motion resulting in a specific asymmetric diffusion.
We performed numerical simulations of this process  and based on the obtained results   argued that the analog HR
can be detected indirectly through the measurement of the dark soliton diffusion.

Our scheme is complementary to the   previous proposals  (see works~\cite{Carusotto,Balbinot,Recati,Larre,Steinhauer1}) that are based
on the measurements   of the correlation between a pair of Hawking quanta.
The method proposed and developed here may be extended to study the stochastic dynamics of other
nonlinear waves~\cite{Konotop} and other analogue gravity problems~\cite{Jaskula,Hung,Fischer,Uhlmann,Fedichev0,Barcelo,Weinfurtner0,Finazzi,Fialko}.

%%%%%%%%%%%%%%%%%%%%%%%%%%%%%%%%%%%%%%%%%%%%%%%%%%%%%%%%%%%%%%

\acknowledgments This work was supported by the National Natural Science Foundation of China (NSFC) under Grants Nos.~11474099 and ~11475063, National Key Research and Development Program of China under Grant No. 2017YFA0304201, Shanghai Program of Shanghai Academic/Technology Research Leader under Grant No. 17XD1401500, and the US National Science Foundation under Grant No. PHY-1620039.

%%%%%%%%%%%%%%%%%%%%%%%%%%%%%%%%%%%%%%%%%%%%%%%%%%%%%%%%%%%%%%%%%%
%\appendix
%\section{}\label{AppendixA}

%%%%%%%%%%%%%%%%%%%%%%%%%%%%%%%%%%%%%%%%%%%%%%%%%%%%%%%%%%%%%%%%%%

%%%%%%%%%%%%%%%%%%%%%%%%%%%%%%%%%%%%%%%%%%%%%%%%%%%%%%%%%%%%%%%%%


\begin{thebibliography}{99}
\bibitem{note100}Since temperature $T$ is very low, only long wavelength phonons can be significantly excited in the system.
\end{thebibliography}
\end{document}